\documentclass[preprint]{aastex}

\shorttitle{Finding Extreme Subdwarfs}
\shortauthors{Marshall}
\usepackage{lscape}

\begin{document}

\title{Finding Extreme Subdwarfs}

\author{J. L. Marshall}
\affil{Observatories of the Carnegie Institute of Washington}
\affil{813 Santa Barbara Street, Pasadena, CA 91101 \\
email: marshall@ociw.edu}


\begin{abstract}

I develop a new technique to identify M-type extreme subdwarfs (esdMs) and 
demonstrate that it is substantially more efficient than previous
methods.  I begin by obtaining spectroscopy and improved photometry
of a sample of 54 late-type halo candidates using the rNLTT reduced
proper motion (RPM) diagram.  From spectroscopy, I find that four of these are 
esdMs, three of which were previously unknown.  From the
improved photometry, I show that all four lie in a narrow RPM corridor that
contains only 4 non-esdMs.  Hence, with good photometry (i.e., without
spectroscopy), it appears possible to select esdM candidates with a 50\%
esdM yield.  This is more than an order of magnitude more efficient than
previous methods.

\end{abstract}




\section{Introduction}\label{sec:intro}

Extreme subdwarfs are very metal-poor red dwarf stars that
are among the oldest stars in the Galaxy.  These objects are generally members of 
the Galactic halo and hence comprise the majority of nearby very metal-poor 
halo stars.  
While extremely metal-poor subdwarfs may be found at a range of spectral types, 
the term ``extreme subdwarf'' generally refers to a late-type (K or M) 
very metal-poor ([Fe/H]$\sim-$2) star.

Extreme subdwarfs are a largely untapped resource in Galactic studies, 
but might prove to be an interesting new population with which to conduct detailed studies
of the local Galactic halo.  
In particular, they may be an interesting population in which to search for the 
most metal-poor stars.
Due to their low luminosity, few extreme subdwarfs are currently known; 
as such, these stars are rarely the focus of Galactic studies.
Nonetheless, with large photometric surveys from which to select candidates 
imminent and  with
large telescopes available for spectroscopic follow-up, 
studies of these nearby but faint halo stars will be feasible.  

Relatively few extreme subdwarfs are currently known, the majority 
having been found in studies targeting nearby stars
\citep[e.g.][]{pCNS3, pmsu1, pmsu2, gizis97}.  
The definition of the M, sdM, and esdM spectral classes by \cite{gizis97} 
includes only 17 esdM stars out of 79 K- and M-dwarfs in the sample.  
\cite{rg05} state that only 13\% of their 367 proper motion-selected late-type dwarfs
are extreme subdwarfs.  
More work has been done recently on targeted searches for ultracool extreme subwarfs, 
i.e., stars with [Fe/H]$\sim-$2 and spectral types later than esdM7.  
These stars are even more elusive, with only a handful currently known.  
A recent census of ultracool subdwarfs is given by \cite{lrs2007}.  

This paper describes a promising technique to find efficiently a significant 
number of bright, nearby extreme subdwarfs.  I select 54
metal-poor subdwarf candidates from a reduced proper motion (RPM) diagram
and obtain spectroscopy and improved photometry from this sample.  I use
spectral line indices to classify the stars into dwarf, subdwarf, 
and extreme subdwarf luminosity classes and then show that the improved
photometry allows definition of a narrow region on the RPM
diagram in which 50\% of the sample are confirmed M-type extreme subdwarfs.  


\section{Sample Selection}\label{sec:sample}

The stars studied in this paper have been selected from the revised NLTT (rNLTT) catalog 
of \cite{gs03} and \cite{sg03}.  The rNLTT matches Luyten's 
original NLTT high-proper motion catalog with the Two Micron All Sky Survey 
\citep[2MASS;][]{2mass} second data release.  
The revised catalog contains 35,725 NLTT stars and 
is relatively complete ($\sim$97\%) for stars brighter than $V<$18 mag with $\delta>-$33$^o$, 
over a region corresponding to the 
second incremental 2MASS data release ($\sim$44\% of the sky).
Because the rNLTT gives proper motions as well as optical and infrared colors, 
it is an ideal resource from which to select a sample of nearby halo, and hence metal-poor, 
stars \citep[][]{sg02}.  

Subdwarfs were selected using a RPM diagram, which plots the reduced proper 
motion, $H_M = m+5 log \mu + 5$,
versus a color index and is used to separate stars into distinct luminosity classes.
This work follows \cite{sg02,sg03} and employs a wide wavelength baseline $(V-J)$ to 
sort high proper motion stars in the rNLTT cleanly into main sequence star, subdwarf, and 
white dwarf classes.  

Subdwarfs are selected from the RPM diagram 
as in \cite{paper1} using discriminator lines as defined by \cite{sg03}.  
Unlike \cite{paper1}, however, this sample was selected from the 
entire rNLTT, i.e., the stars were not required to be equatorial.  
The stars studied here are required to have $(V-J)>$3.5 mag.  
It should be noted that this criterion yields a sample of very 
late-type metal-poor (halo) stars, though they may not necessarily be extreme subdwarfs.
This color cut was employed simply because the rNLTT photometry 
is not accurate enough to select reliably only the bluer stars 
in this region, where the more metal-poor extreme subdwarfs are expected to lie.  
This is discussed in more detail in Section~\ref{sec:disc}.

\begin{figure}
\plotone{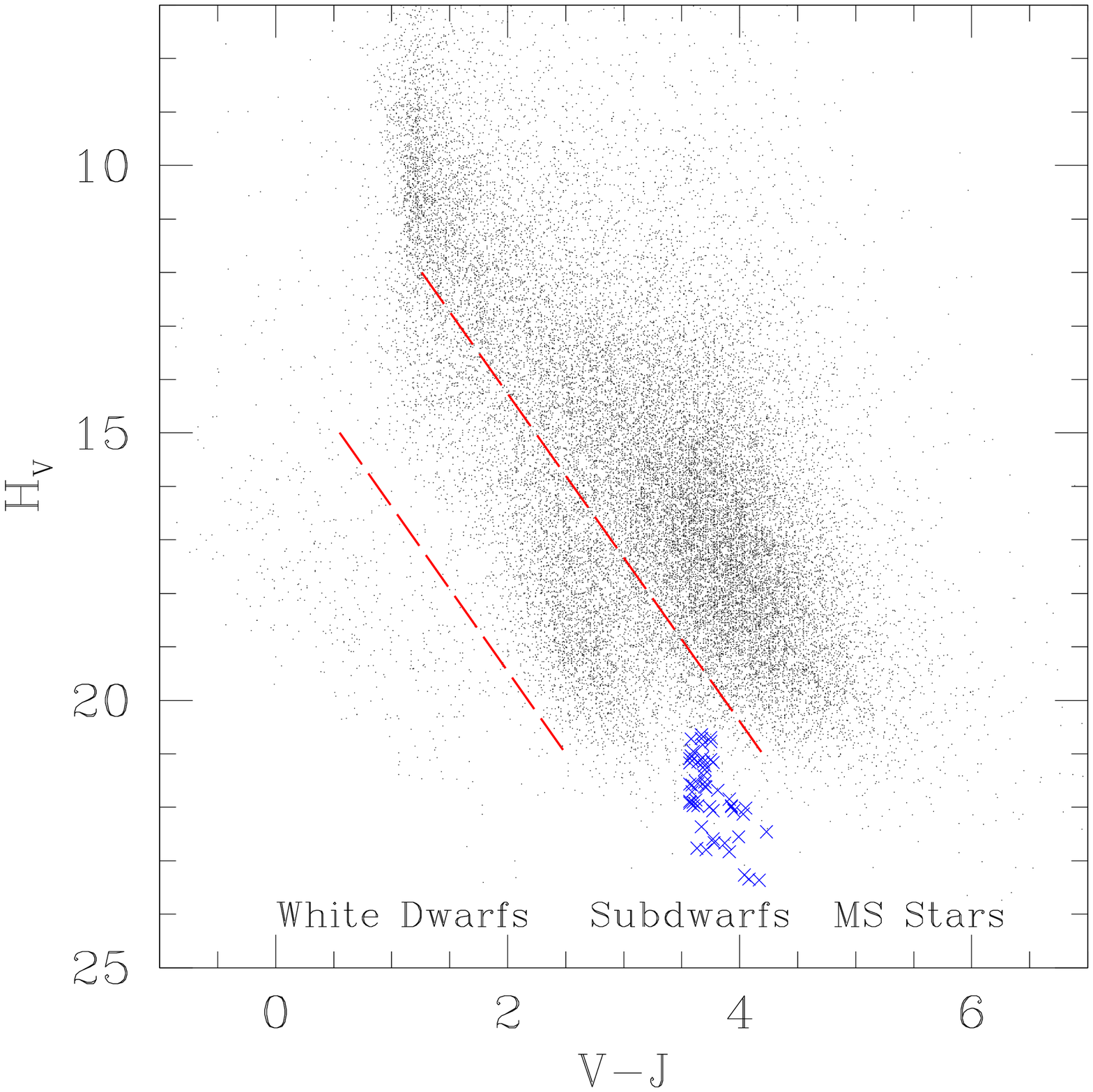}
\figcaption{
Reduced proper motion diagram of the selected extreme subdwarfs.  
This RPM diagram is constructed using the USNO-A 
photometry from the rNLTT.  The diagram shows all of the 35,725 stars in the rNLTT; 
the crosses are the 54 subdwarfs selected for this study.
\label{fig:rpmold}
}
\end{figure}

The above criteria yield 54 subdwarfs.  These stars are marked on the 
RPM diagram of the entire rNLTT in Figure~\ref{fig:rpmold}.


\section{Observations and Data Reduction}\label{sec:data}

Photometric and spectrophotometric observations were obtained during 2003--2004 at the MDM
Observatory on Kitt Peak, Arizona, and at the Cerro Tololo Inter-American
Observatory (CTIO) in Chile. 

\subsection{Photometry}

Photometry was obtained at MDM Observatory at the 1.3m telescope using the 1024x1024
``Templeton'' CCD detector.  At the CTIO 0.9m telescope, the 2048x2046 CCD detector was used.
Standard BVRI Johnson/Kron-Cousins filters were used for all observations.  
Photometric standard star fields were observed often during each night, 
generally one standard field per hour.  
An effort was made to observe photometric standard stars with red colors 
in order to match the colors of the very red target stars.  
Unfortunately, there are very few standard stars with colors as red as the objects studied here, 
and the extrapolation of the photometric color term to very red colors 
may have some bias.  Nonetheless, the photometry presented here is self-consistent across 
the entire observing period.  The photometric data acquisition and reduction 
procedure is described in greater detail by \cite{paper1}; the photometric data presented here 
was obtained during the same nights as those presented by \cite{paper1}.  

\subsection{Spectrophotometry}

Spectrophotometric observations were obtained with the CTIO 1.5m telescope
and with the MDM Observatory 2.4m telescope.
Observations generally covered the wavelength range 6500--8100\AA. 
All observations were obtained using a
North--South oriented slit, and observations were made as the targets transited
the meridian ($\pm$ 1 hour of transit) in order to minimize slit losses. 

The CCDS Spectrograph \footnote{See http://www-astronomy.mps.ohio-state.edu/MDM/CCDS/} 
was used for all observations
obtained at the MDM 2.4m telescope. The 350 l/mm grating and a 1\farcs0 
(87 micron) slit was used for all observations, producing a spectral resolution
of 3.4 \AA \ per resolution element.  

The RC Spectrograph \footnote{See http:
//www.ctio.noao.edu/spectrographs/60spec/60spec.html.} was used to obtain
observations at the CTIO 1.5m telescope. Grating ``35/I''
was used at 20.8 degrees tilt to obtain the appropriate wavelength range. A 2\farcs0 
(110 $\mu$m) slit provided similar resolution to those obtained at MDM.  

Spectrophotometric observations of 42 stars were obtained.  
The final spectra were assembled by averaging multiple (typically 4--6) 
900-second observations of the target in order to achieve the desired 
signal-to-noise ratio ($S/N$) of $\sim$10 (although some observations 
have lower $S/N$ due to poor conditions). 
Spectrophotometric standard stars were also observed on each night, generally
about once every 2--3 hours throughout the night.  

Spectra were reduced using IRAF\footnote[1]{IRAF is distributed by
the National Optical Astronomy Observatory, which is operated by the Association of
Universities for Research in Astronomy, Inc., under cooperative agreement with the
National Science Foundation.}, using standard data reduction routines
to bias subtract, flatten, and extract the spectra.
Spectra were extracted from each individual frame, then coadded together to form 
the final spectrum of the target.  


\section{Results}\label{sec:results}

\subsection{$BVRI$ Photometry}

Table~\ref{table:phot} presents new photometry for the 54 selected stars, 
along with existing 2MASS photometry.  
Column 1 of Table~\ref{table:phot} gives the NLTT identifier of each target; 
columns 2--3 give the position of the star.  
Columns 4--7 present the $V$, $B-V$, $V-R$, and $V-I$ photometry measured in this work.  
The $V-J$, $V-H$, and $V-K$ color indices in columns 8--10 are 
formed by simply taking the difference between the $V$-band color measured here 
and the infrared 2MASS photometry.  
Column 11 indicates how many photometric measurements were averaged together to form the 
final photometry.  
In general, multiple photometric measurements were obtained of each star:
128 measurements of 54 stars were obtained, and 43 stars have more than one measurement.

Figure~\ref{fig:rpmnew} shows a RPM diagram constructed with the new photometry presented here, 
along with improved photometry of 564 candidate subdwarfs from \cite{paper1}.
Note that two of the stars in this sample have scattered to lower 
reduced proper motions; this is most likely due to a misidentification, 
either in this work or in the rNLTT (or the original Luyten catalog).

\subsubsection{Photometric Errors}

\begin{figure}
\plotone{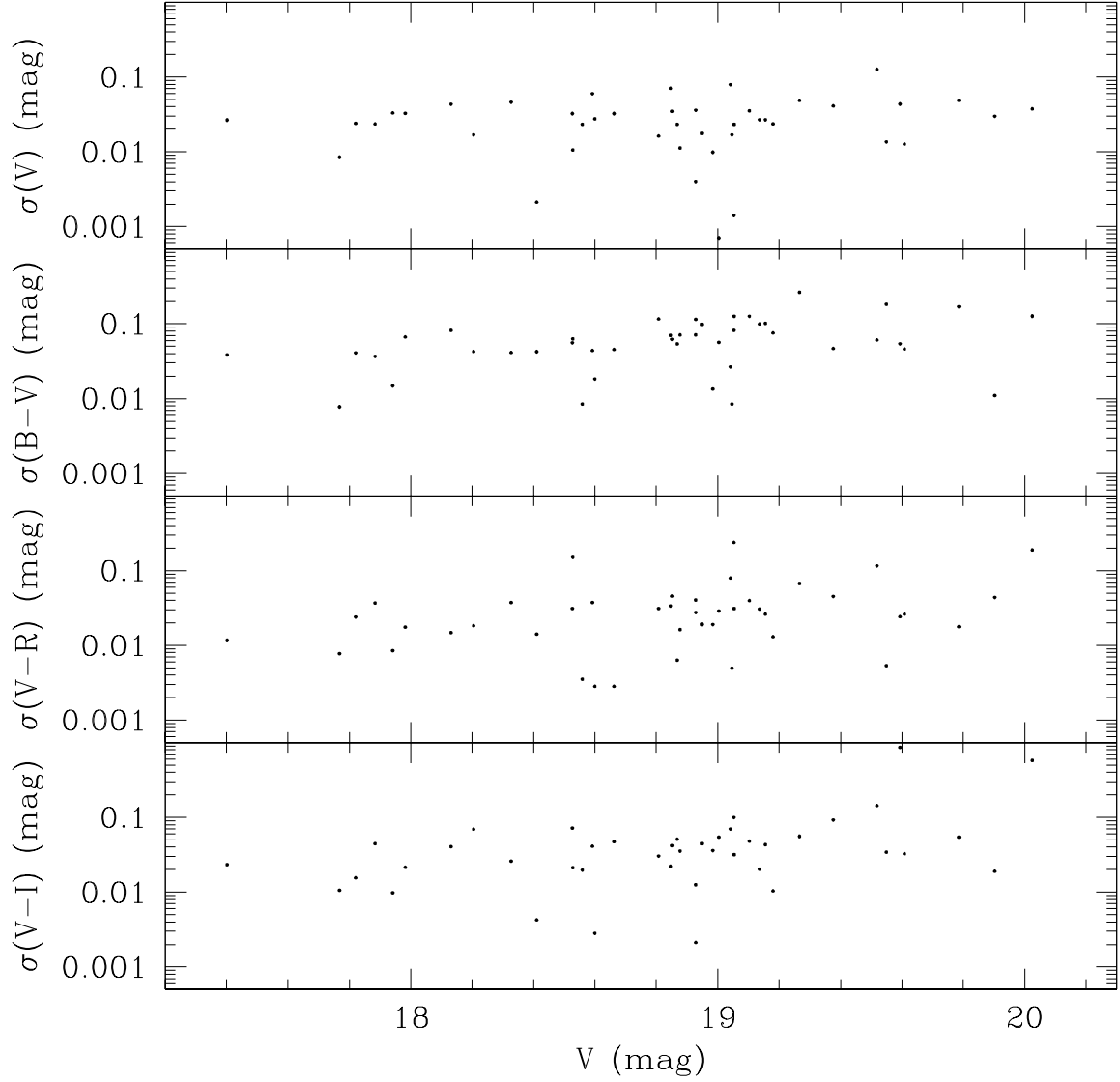}
\figcaption{
Standard deviation of multiple photometric measurements of a given star as a function of magnitude.  
The mean photometric error in these measurements is 0.03 mag in $V$, 0.07 mag in $B-V$, 
0.04 mag in $V-R$, and 0.07 mag in $V-I$.
All of this error is likely to be due to measurement error, as the stars are not expected to be
variable.
\label{fig:stdevall}
}
\end{figure}

The photometric errors were determined by comparing the photometry of each of the
43 stars with multiple measurements.  See Figure~\ref{fig:stdevall}.
The error estimates, (0.03, 0.07, 0.04, 0.07) mag in 
($V$, $B-V$, $V-R$, $V-I$) were derived from
the average of the standard deviations of these multiple measurements.
(See \cite{paper1} for a more thorough discussion.)

\begin{figure}
\plotone{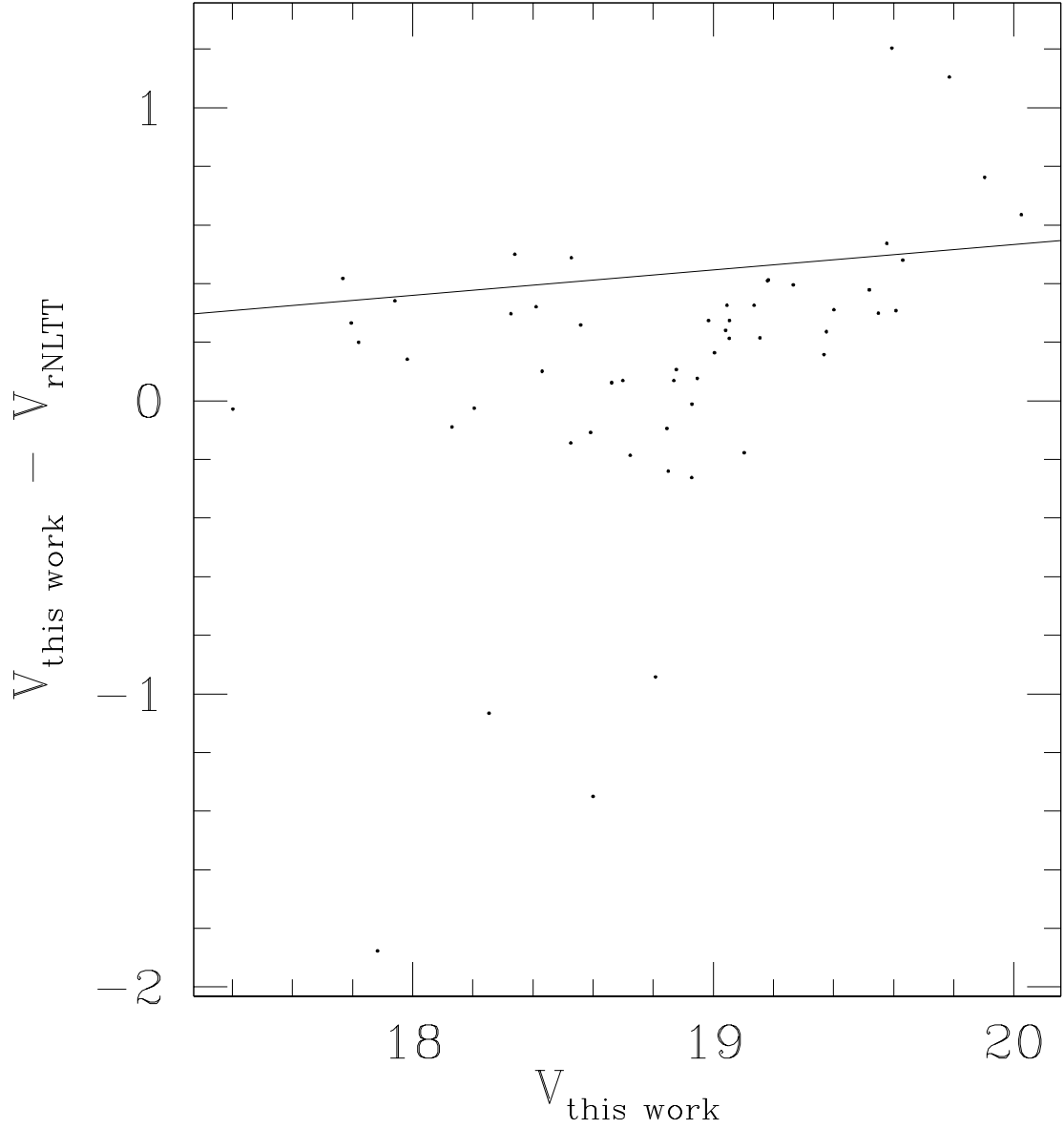}
\figcaption{
Comparison of the photometry derived here and that given by the rNLTT.  
The relation shows a fit to the typical difference between the photometry 
derived by \cite{paper1} and the rNLTT photometry described by $V_{this work} - V_{USNO-A} =
0.012 + 0.087 (V_{this work} - 14)$.  If 6 outliers are excluded, the
scatter in these data is about 0.25 mag, in good agreement with \cite{sg03}.
\label{fig:comprnltt}
}
\end{figure}

\subsubsection{Comparison with rNLTT Photometry}

rNLTT photometry \citep[derived from USNO-A;][]{monet96,monet98} is quite inaccurate at magnitudes 
as faint as those studied here.  
Figure~\ref{fig:comprnltt} compares the photometry measured in this work with 
that given by rNLTT.  
Overplotted on the figure is the relation found by \cite{paper1} describing 
the comparison between modern, more accurate photometry and the rNLTT values.  
If 6 outliers are removed, the scatter is about of 0.25 mag, in good agreement
with the estimate by \citet{sg03}.
These may be due to a misidentification or perhaps 
simply to poor photometry in the rNLTT.

\subsection{Spectroscopic Classification}\label{sec:lum}

The stars are classified following \cite{gizis97}.
This technique defines a set of spectral line indices to be applied to moderate-resolution 
(R$\sim$3000) spectra to determine whether a given star is a dwarf ([m/H]$\sim$0.0), 
subdwarf ([m/H]$\sim$-1.2), or extreme subdwarf ([m/H]$\sim-$2.0).
The line indices are centered on the CaH and TiO molecular features 
that are prevalent and easily measurable in moderate resolution spectra of 
late-type stars and whose strength is dependent on the star's metallicity.
The CaH indices are compared to the TiO index and discriminator lines are drawn to
separate the three classes of stars. 

\subsubsection{Measurement of Line Indices}\label{sec:measind}

The line indices described by \cite{gizis97} are measured by integrating 
over a spectral region containing the absorption feature of interest, then 
comparing the result to a nearby ``continuum'' region.  Several indices are identified 
by \cite{gizis97} as being indicative of a star's metallicity; 
these line indices are defined in Table~\ref{table:gizisind}.

Line indices are measured using the IRAF utility ``sbands'', 
which simply integrates the flux in the spectrum over the bandpass of interest.  
Indices are then constructed via the following relation, using the TiO5 index as an example:
\begin{equation}
{\rm TiO5} = \frac{ \int_{7126}^{7135} F _\lambda d\lambda / 9 \AA }{ \int_{7042}^{7046} F _\lambda d\lambda / 4 \AA }
\end{equation}

\subsubsection{Determination of Spectroscopic Classification}

\begin{figure}
\plotone{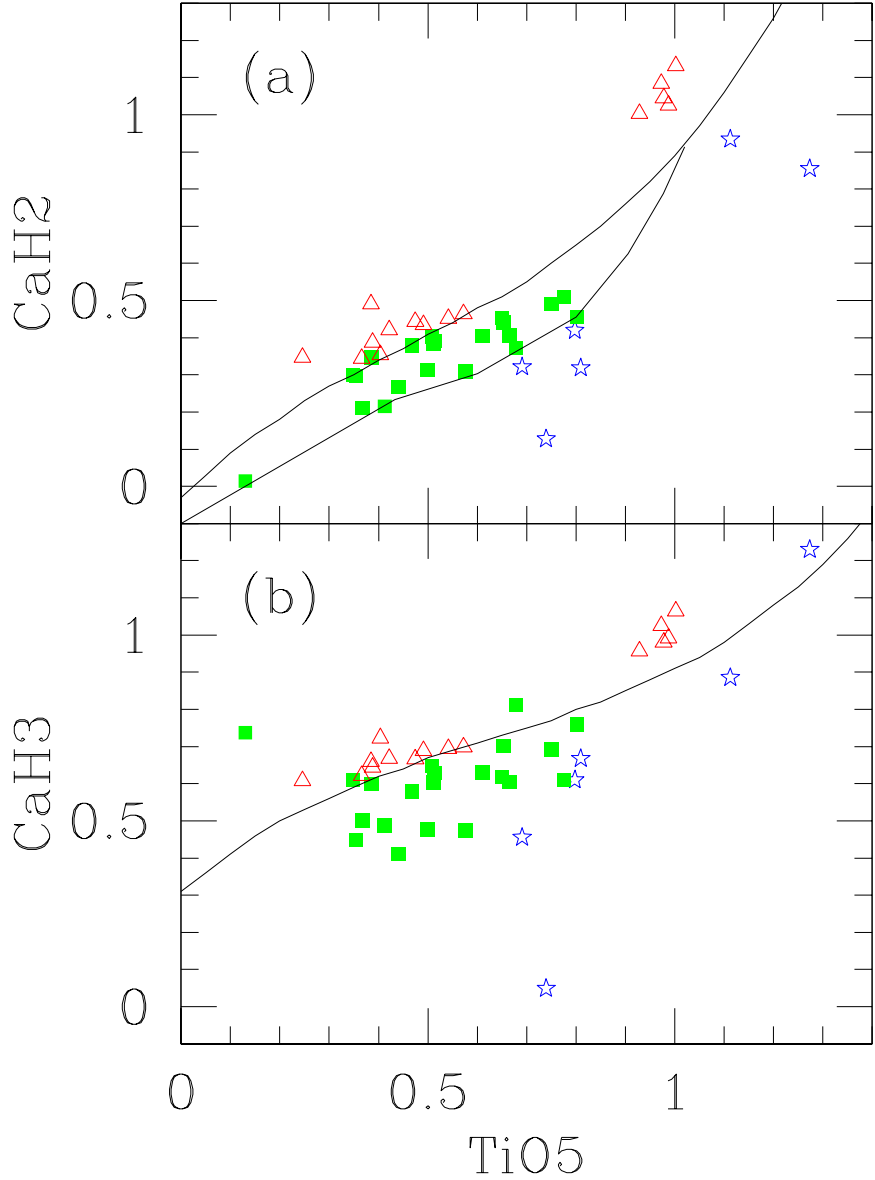}
\figcaption{
Classification scheme for M-dwarfs, subdwarfs, and extreme subdwarfs as defined by \cite{gizis97}.  
Spectral line indices are measured in each spectrum and have been calibrated in order 
to divide the stars by luminosity class.  In the above figure M-stars are open 
triangles, subdwarfs are filled squares, and extreme subdwarfs are represented by 
open stars.  
(a) plots the main criterion, the CaH2/TiO5 ratio, that divides the 
stars into the three luminosity classes, 
using discriminators as defined by \cite{lrs2003} \citep[based on][indices]{gizis97}.
(b) provides a secondary confirmation of the classification scheme, separating 
dwarfs from the subdwarfs.
\label{fig:cah123pts}
}
\end{figure}

Spectroscopic classification is determined as described by \cite{gizis97}.  
Following the procedure described therein, the CaH2 and CaH3 indices are compared 
to the TiO5 index for each star, and discriminator lines are drawn to distinguish 
the luminosity classes.  Figure~\ref{fig:cah123pts} shows the classification scheme
as defined by \cite{lrs2003}, based on the \cite{gizis97} indices:
Figure~\ref{fig:cah123pts}(a) separates dwarfs (top, open triangles) 
from subdwarfs (middle, filled squares) and extreme
subdwarfs (lower right, open stars).  The discriminator in Figure~\ref{fig:cah123pts}(b)
separates dwarfs (top) from subdwarfs and extreme subdwarfs (bottom).

Once the stars have been classified according to luminosity class, 
the relations given in \cite{gizis97} are used to determine the spectral type of each star.  
These relations are stated to be accurate to $\pm$0.5 subclass.

Table~\ref{table:gizis} presents the indices measured in each spectrum: 
Column 1 gives the NLTT identifier and columns 2--3 are the new $V$ and $V-J$ photometry.  
Columns 4--5 give the date and telescope on which the spectrum was obtained.
The indices measured in each spectrum are given in columns 6--8.
Column 9 of Table~\ref{table:gizis} gives the classification of each star as determined in this work.
Spectral classifications are given to 0.5 subclass for M-type dwarfs, subdwarfs, and 
extreme subdwarfs.  For earlier K-type dwarfs, the relations are somewhat less accurate and 
these determinations are given to the nearest subclass, or if found to be earlier than 
K5, simply cited as ``K''.

For three stars (NLTT37223, NLTT50476, and NLTT55103) more than one spectrum was obtained.  
In these cases, the indices were measured for each spectrum, and they agree relatively well.  
The three spectra of NLTT50476 yield spectral types of sdM4.5, M4.0, and sdM4.5.  
NLTT55103 has two spectra which give spectral types of sdM5.0 and sdM4.5.
The spectral type of sdM4.5 is adopted for both NLTT50476 and NLTT55103.
The two spectra of the extreme subdwarf with multiple spectra, NLTT37223, 
do not agree quite as well; the spectra give spectral types of esdM3.5 and esdM6.0.  
Due to the discrepancy between the spectral types, a final spectral type for this star
is not assigned, although it is noted that the spectrum which yields a type of esdM6.0
is significantly more noisy, which may have influenced the spectral determination.  

\begin{figure}
\plotone{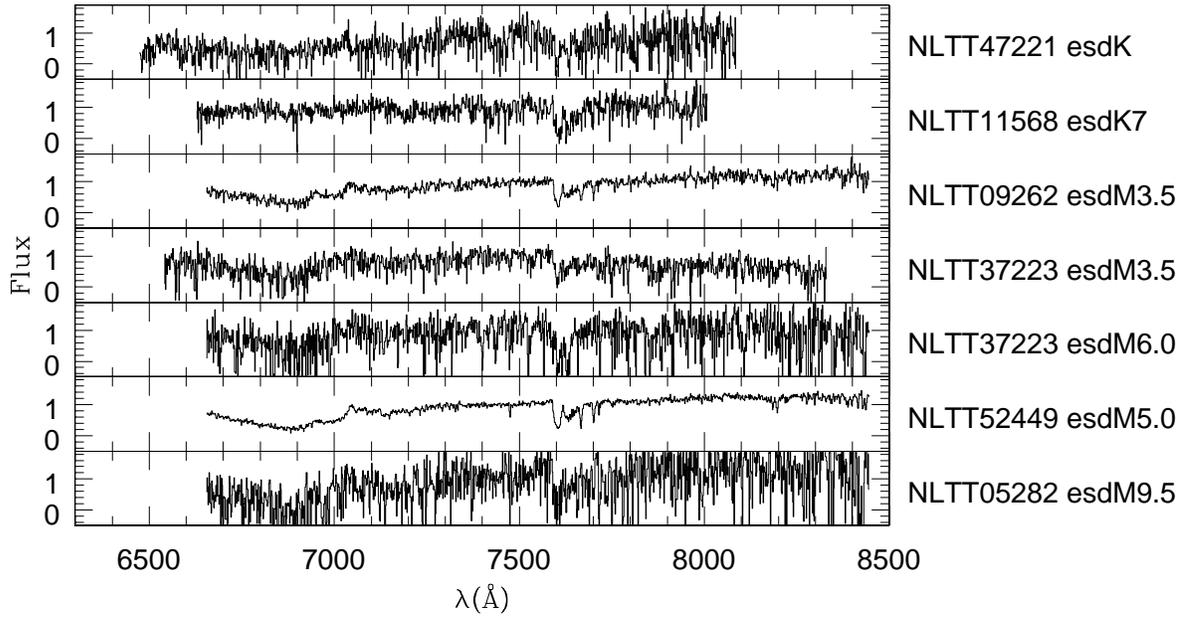}
\figcaption{
Spectra of the 6 extreme subdwarfs in the sample.
Note that two spectra (with different derived spectral types) are shown for NLTT37223.
The spectral regions near 6867 and 7594 \AA\ are somewhat contaminated by atmospheric 
absorption.  
\label{fig:spec_esd}
}
\end{figure}

\begin{figure}
\plotone{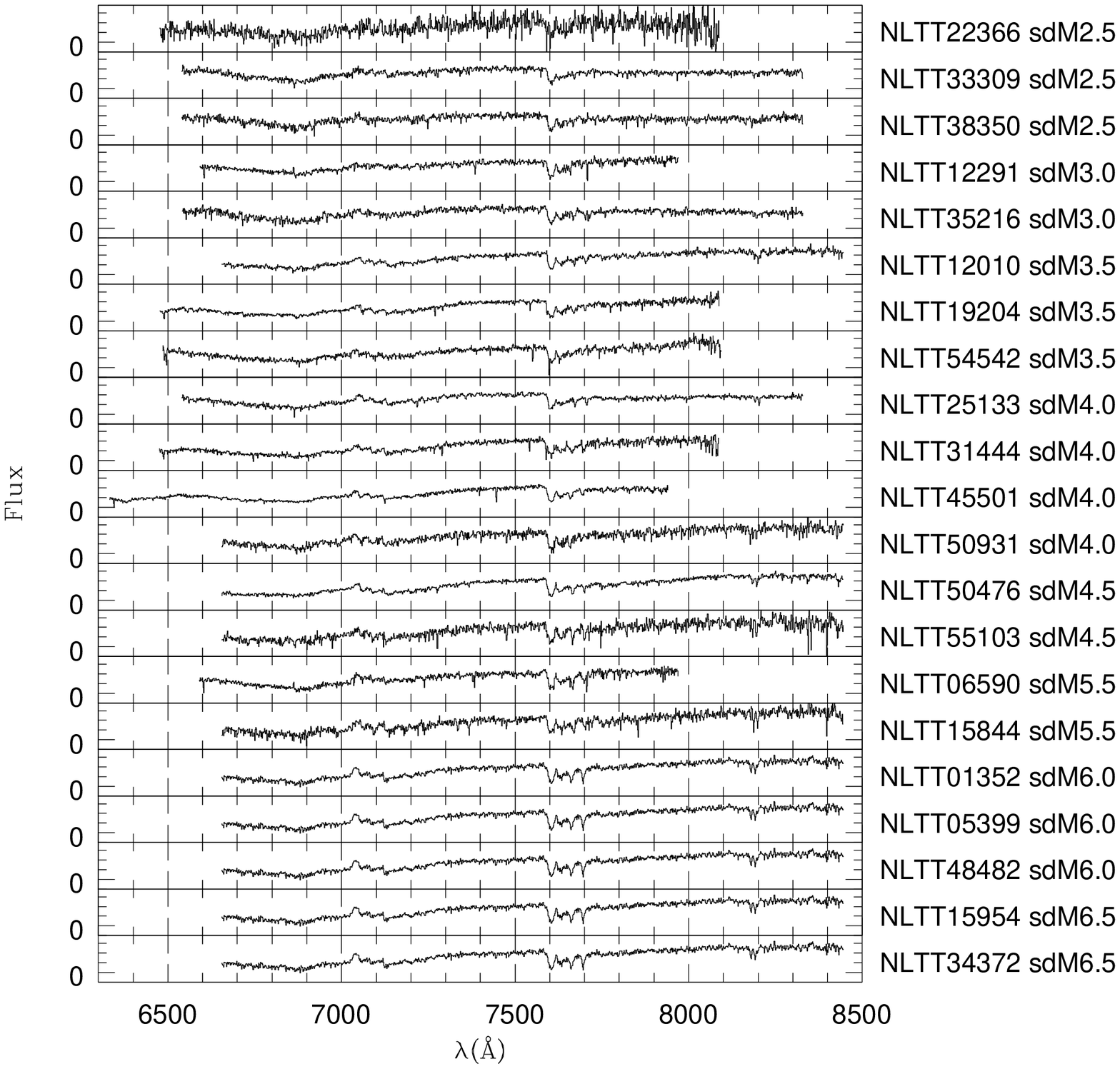}
\figcaption{
Spectra of the 21 subdwarfs in the sample.
The spectral regions near 6867 and 7594 \AA\ are somewhat contaminated by atmospheric 
absorption.  
\label{fig:spec_sd}
}
\end{figure}

\begin{figure}
\plotone{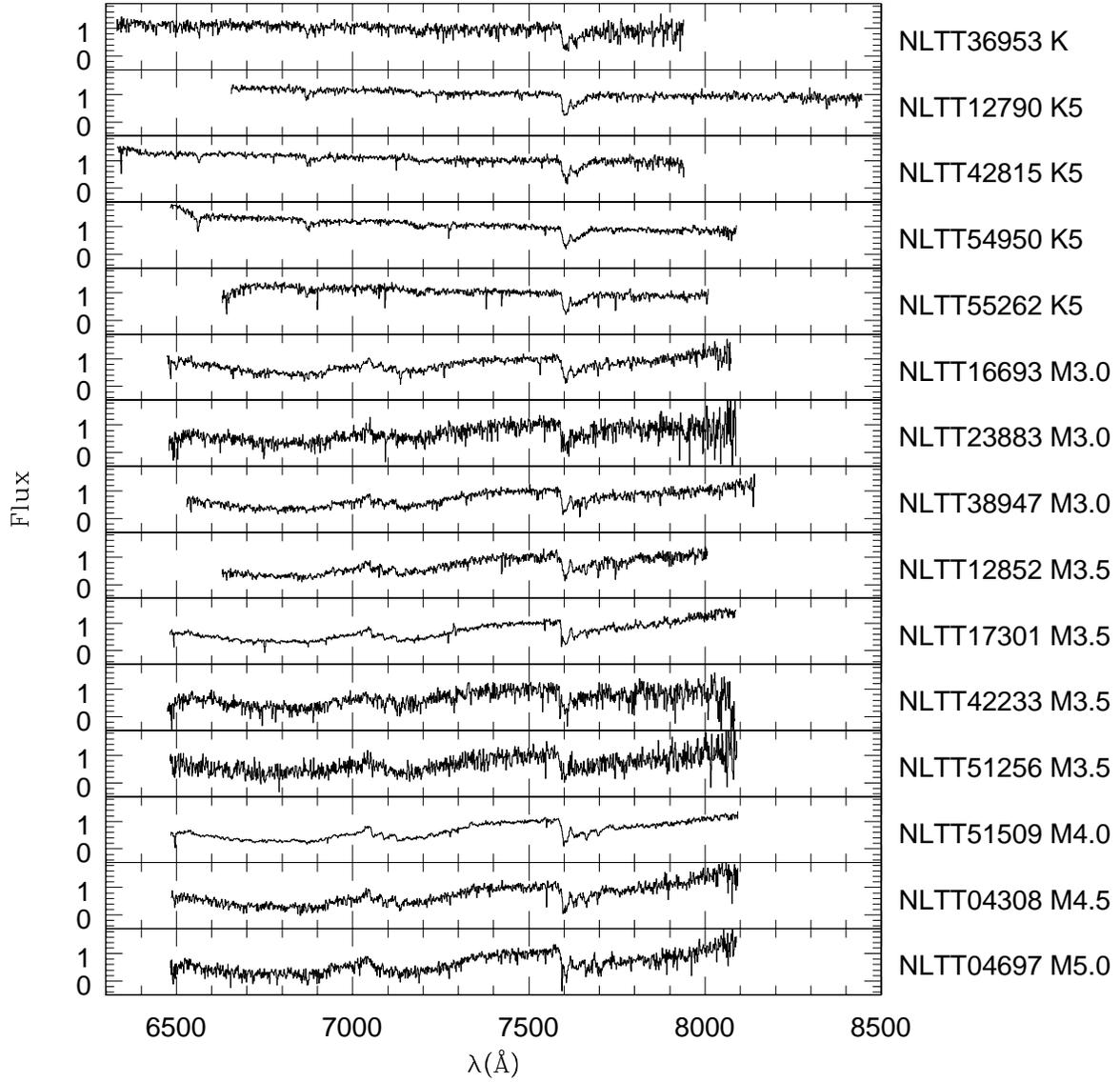}
\figcaption{
Spectra of the 15 K and M dwarfs in the sample.
The spectral regions near 6867 and 7594 \AA\ are somewhat contaminated by atmospheric 
absorption.  
\label{fig:spec_m}
}
\end{figure}

Figures~\ref{fig:spec_esd}, \ref{fig:spec_sd}, and \ref{fig:spec_m} show the spectra of the 
6 extreme subdwarfs, 21 subdwarfs, and 15 K- and M- dwarfs studied here.
Note that Figure~\ref{fig:spec_esd} displays both spectra of NLTT37223 
to show the discrepancy between the two spectral type determinations.

\subsubsection{Comparison with Spectral Classifications in the Literature}

Four stars in this sample have previously been classified in the literature; 
all of these previous classifications have been made using the technique described here.  
One extreme subdwarf in this sample has been previously studied: 
\cite{rg05} classify NLTT52449 as esdM5.0, the same classification that is derived here.  
\cite{gr97} determine the spectral types of both NLTT22366 and NLTT25133.  
They classify these stars as sdM3.5 and sdM4.5, while in this work they are 
classified as sdM2.5 and sdM4.0, respectively.
NLTT51509 is one of the original subdwarfs used to establish this technique: 
\cite{gizis97} classifies it as M4.5; here, a spectral type of M4.0 is measured. 
The roughly $\pm$0.5 subclass differences seen in these common determinations are 
in accord with the expected uncertainty in the technique.  


\section{Discussion}\label{sec:disc}


Six extreme subdwarfs are found in this sample of 42 candidates.  
Four of the six are M-type extreme subdwarfs (NLTT09262: esdM3.5, NLTT37223: esdM3.5/6.0,
NLTT52449: esdM5.0), although the exact type of NLTT37223 remains ambiguous.  
The classification of NLTT05282 (esdM9.5) is less secure since it is not clear that 
this technique is applicable to such late spectral types.  
This may be an interesting target for further study.  
Two stars, NLTT47221 and NLTT11568, are found to be esdK stars.  

\begin{figure}
\plotone{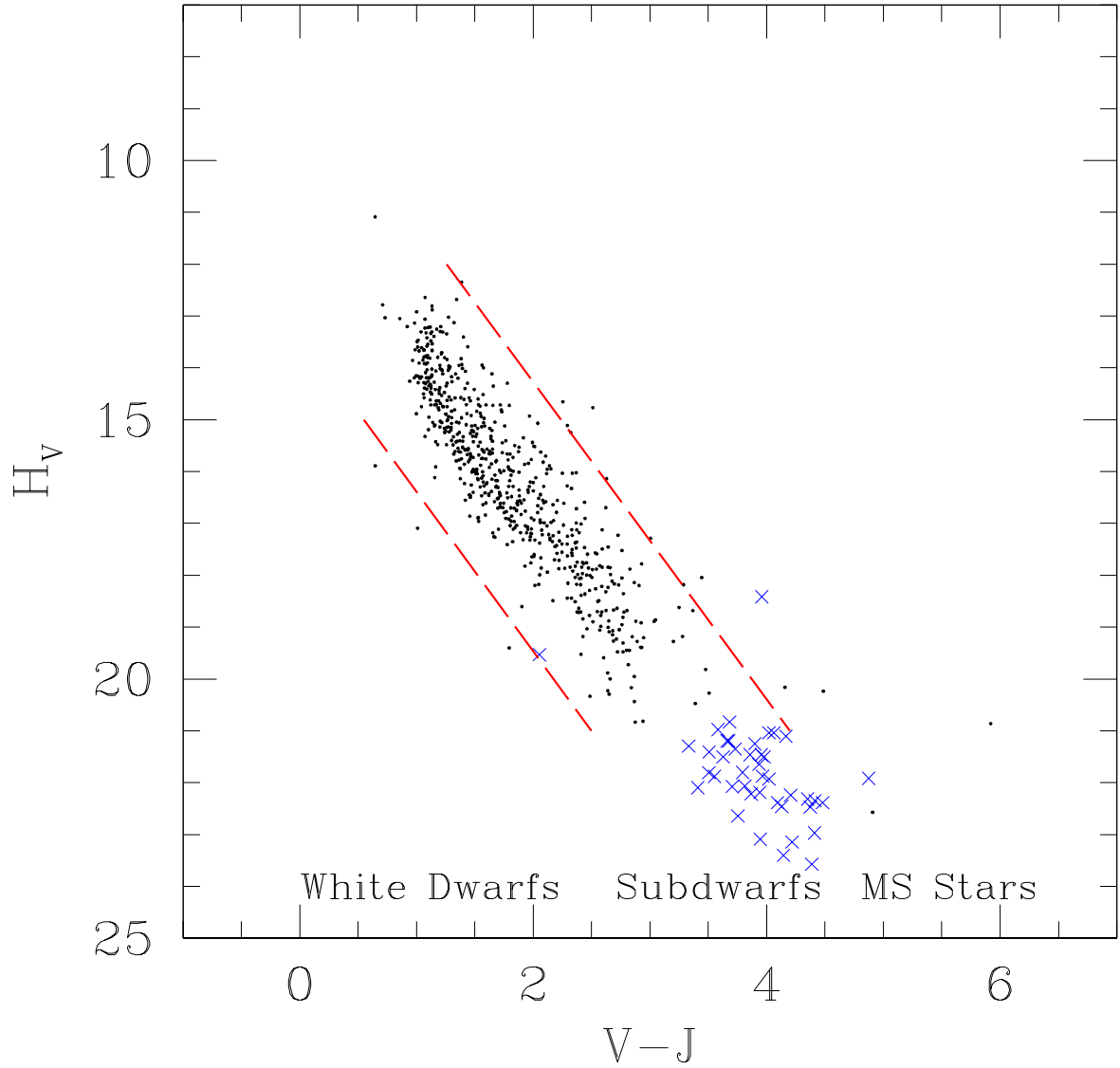}
\figcaption{
Reduced proper motion diagram constructed using the improved photometry.
The small points in this figure show the photometry of the earlier-type (F, G, K) subdwarfs 
studied by \cite{paper1}.  The crosses represent the new photometry of the 
54 late-type subdwarfs studied here.
\label{fig:rpmnew}
}
\end{figure}

\begin{figure}
\plotone{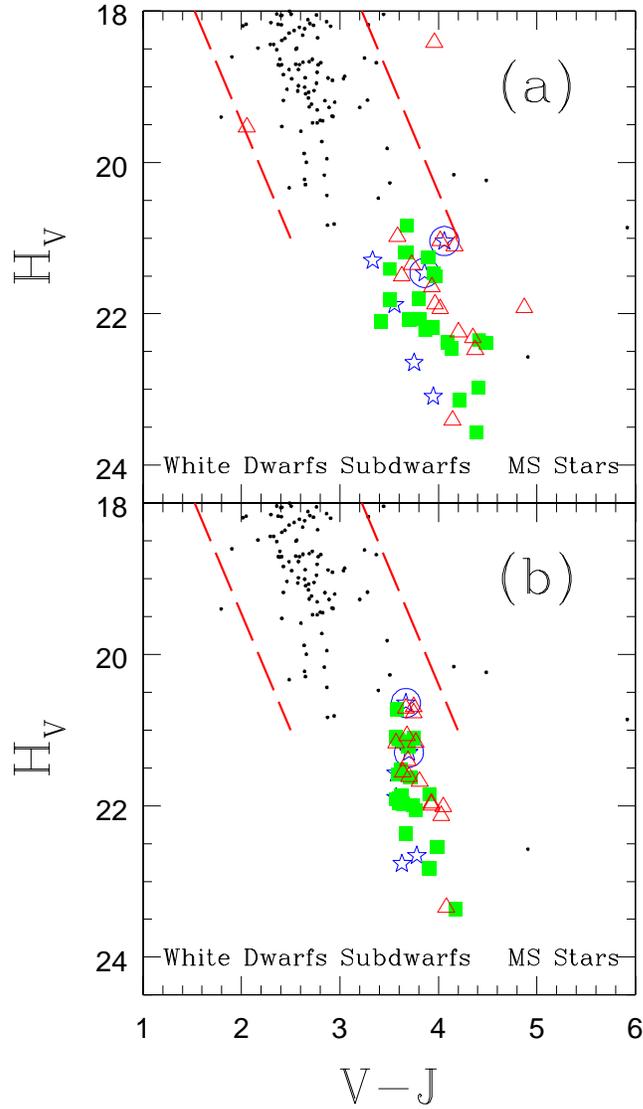}
\figcaption{
(a) The newly classified sources are plotted on the RPM diagram as before.  
Symbols are the same as in Figure~\ref{fig:cah123pts}.
With the improved photometry, the extreme subdwarfs 
lie blue-ward of the subdwarfs and solar-metallicity dwarfs.  
The circled triangles are classified as esdK stars; due to some ambiguity 
in the classification between esdK, sdK and K-dwarf stars, these stars are 
possibly not truly very metal-poor.  
(b) The original RPM diagram using rNLTT photometry (as in Figure~\ref{fig:rpmold})
but with luminosity class indicated for each star.  
Note that there is no obvious separation between the classes.
\label{fig:rpmoldpts}
}
\end{figure}

Figure~\ref{fig:rpmoldpts}(a) shows the RPM diagram presented in Figure~\ref{fig:rpmnew}, 
with the three luminosity classes shown as different symbols.
This figure shows the separation of the three luminosity classes quite distinctly:
Extreme subdwarfs (open stars) lie to the left of the diagram, as expected due to their 
lower metallicity which makes them bluer at a given luminosity.
The subdwarfs (filled squares) are slightly redder than the extreme subdwarfs, and the 
near-solar metallicity M dwarfs (open triangles) lie to the right of the diagram. 
The two circled points are the esdK stars.  Since the classification scheme 
was devised to classify early- to mid-M dwarfs and subdwarfs, it is 
less secure for K-type stars.  It is possible that these two stars classified as 
esdK are in fact sdK or K-dwarfs.  This may also be seen in Figure~\ref{fig:cah123pts}; 
the esdK stars in this figure are the outliers to the top right.

Ten percent (4/42) of the subdwarfs studied spectroscopically are shown to be 
M-type extreme subdwarfs according to this technique.  
This is a relatively high yield, given the difficulty in finding and observing 
candidate extreme subdwarfs. 
An even higher yield could be had if one were to select candidate extreme 
subdwarfs using more accurate photometry, targeting stars only 
from the blue region of a RPM diagram. 
As a demonstration of this point, Figure~\ref{fig:rpmoldpts}(b)
shows the RPM diagram using the original rNLTT photometry but with the 
luminosity classes of the stars indicated as in Figure~\ref{fig:rpmnew}.  
In this Figure there is no obvious way to distinguish the stars, and although 
the extreme subdwarfs lie in or near the newly defined region, it is no longer 
convincing that the entire region preferentially contains extreme subdwarfs.

\begin{figure}
\plotone{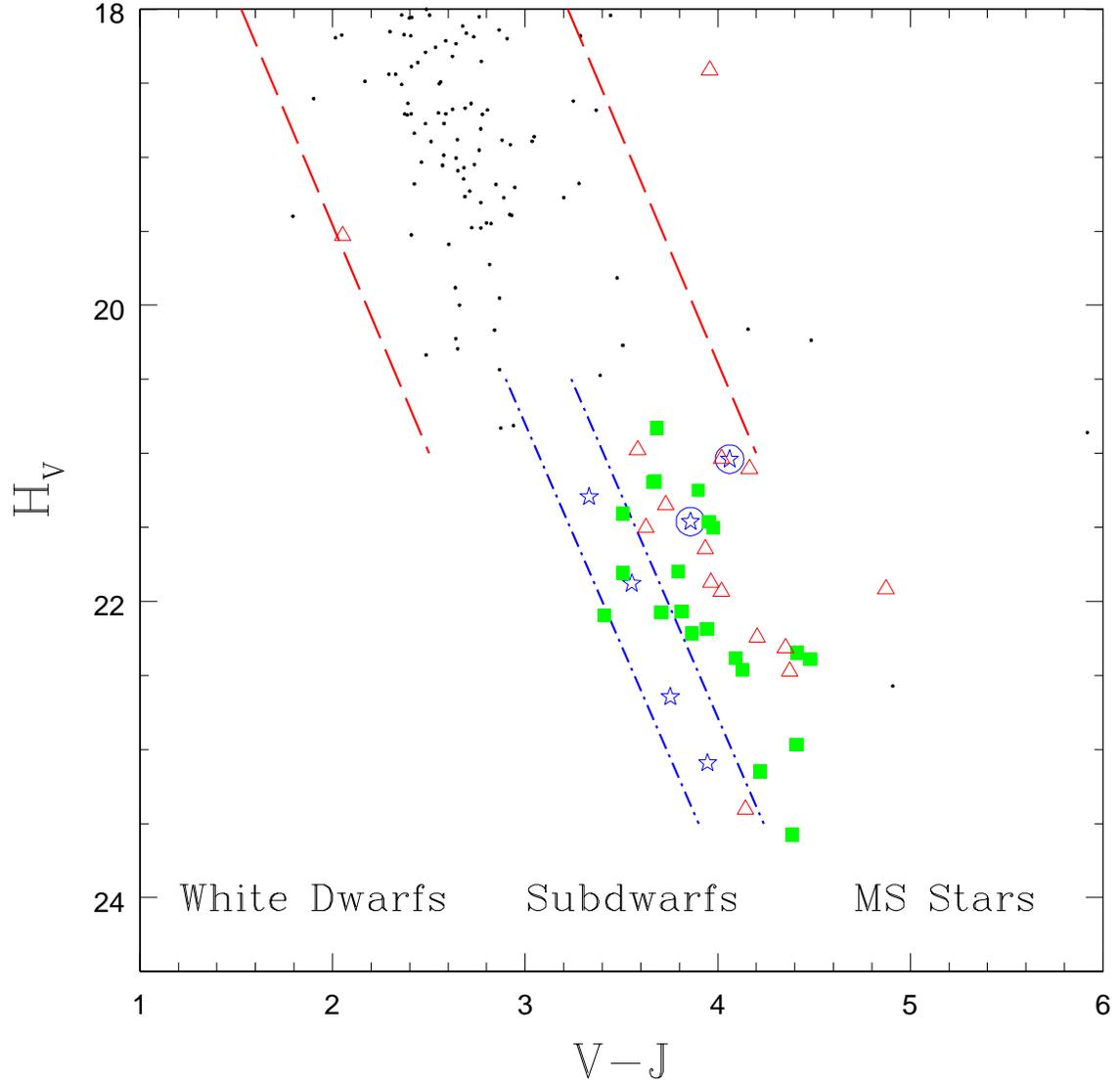}
\figcaption{
New discriminator (dash-dot) lines are drawn to define the late-type extreme subdwarf
region of the RPM diagram; fifty percent of the stars in this region are extreme subdwarfs.
This result could facilitate future searches for extreme subdwarfs 
in databases with accurate photometry.  
\label{fig:rpmnewpts}
}
\end{figure}

As discussed in Section~\ref{sec:sample}, the stars studied here were not particularly selected to be 
extreme subdwarfs, simply to be very late-type subdwarfs, 
i.e., they were selected to lie within the restricted subdwarf discriminators 
of the RPM diagram as in \cite{paper1}.  
However, with the exception of the two esdK stars, the extreme subdwarfs 
lie on a well-defined relation in the RPM diagram.  
A line is fit to these points, and a new set of extreme subdwarf discriminator 
lines is drawn with the equations $H_V = 3\times(V-J) + 10.79$ and $H_V = 3\times(V-J) + 11.79$.
These lines are shown in Figure~\ref{fig:rpmnewpts}.
Fifty percent of the objects within this region are subdwarfs.  
It is somewhat suggestive that all four extreme subdwarfs lie on a rather straight line; 
larger samples of stars with accurate photometry will be needed to refine these 
new discriminator lines.

It is now apparent that, although some
scatter is still present, the more metal-poor extreme subdwarfs lie to the
left of the RPM diagram, while the solar-metallicity dwarfs are towards the right, 
exactly as expected.  
It is somewhat surprising that the M-dwarf sequence lies so close to the subdwarf sequence; this may
be due to a change in the slope of the red end of the M-dwarf main sequence.
This will have to be investigated more carefully by constructing a color-magnitude diagram of
late-type stars with known distances, but this result is opposite that found by \cite{g4588}.
Since the extreme subdwarfs lie to the blue side of the RPM diagram, future searches
for extreme subdwarfs should target only the defined region.
This will be more straightforward when studying stars with accurate photometry at faint magnitudes
(e.g., stars found in the SDSS catalog).

\section{Summary and Future Work}

A new, efficient method for finding extreme subdwarfs has been described.  
New photometry and spectroscopic observations of 54 subdwarfs 
has been presented, allowing for the construction of an accurate RPM diagram 
of the targets.  Using spectral line indices measured in the moderate resolution 
spectra and a classification technique defined by \cite{gizis97}, 
42 late-type dwarfs are classified and 4 esdM stars are found, 
3 of which are previously unclassified.  
A new selection method is defined which uses a RPM diagram 
to select more efficiently a sample of extreme subdwarfs.
With the new selection criteria, $\sim$50\% of candidates should be confirmed extreme subdwarfs.  
It should be noted that this technique has a strong kinematic bias
inherent in the sample selection and should not be used to define an unbiased sample 
of extreme subdwarfs.  Rather, this technique selects stars on the basis of their 
high proper motion and is therefore only sensitive to stars with velocities perpendicular
to the line of sight.  

Although the entire rNLTT has now been mined for extreme subdwarfs, there are 
prospects for finding even more nearby extreme subdwarfs.  Larger proper 
motion surveys now exist \citep[e.g.][]{lspm}.  Furthermore, 
the Sloan Digital Sky Survey (SDSS) SEGUE project contains proper motion 
information for many more Galactic stars than was previously available.  
Applying the techniques developed here to these larger databases will certainly produce 
many more candidate extreme subdwarfs for follow-up spectroscopic study.  

It should be noted that the rNLTT is not complete for magnitudes as faint as the 
stars studied here.  Searching for these stars in a catalog with a fainter 
brightness limit is sure to produce more candidates, although fainter targets will be more 
difficult to follow-up spectroscopically.  
With large, efficient telescopes, however, even very faint extreme subdwarfs 
will be accessible for follow-up study.  


\acknowledgements

The author acknowledges many useful conversations about the sample selection with A. Gould 
and many enjoyable observing runs with D. L. DePoy.


%
%
\newpage
\clearpage

%
%
\newpage

\begin{deluxetable}{rccrrrrrrrc}
\tablecolumns{11}
\tablewidth{0pc}
\tablecaption{
Photometry
\label{table:phot}
}
\tablehead{
\colhead{NLTT} & \colhead{$\alpha$(2000)} & \colhead{$\delta$(2000)} & \colhead{$V$} & \colhead{$B-V$} & \colhead{$V-R$} & \colhead{$V-I$} & \colhead{$V-J$} & \colhead{$V-H$} & \colhead{$V-K$} & \colhead{n} 
}
\startdata

 1352 & 00:25:51.2 & $-$07:48:08 & 19.053 & 2.079 & 1.169 & 2.945 & 4.386 & 4.813 & 5.122 & 2 \\
 4308 & 01:18:08.4 & +49:57:37 & 19.180 & 1.693 & 1.233 & 2.808 & 4.164 & 4.629 & 4.970 & 3 \\
 4697 & 01:25:10.2 & +46:10:27 & 19.594 & 1.658 & 1.443 & 2.804 & 4.873 & 5.389 & 5.585 & 3 \\
 5282 & 01:34:59.2 & $-$09:04:42 & 19.136 & 1.840 & 1.211 & 2.594 & 3.947 & 4.343 & 4.331 & 3 \\
 5399 & 01:36:57.1 & $-$01:02:20 & 19.903 & 1.612 & 1.315 & 2.988 & 4.480 & 5.054 & 5.376 & 3 \\
 6590 & 01:58:28.2 & +17:01:04 & 18.947 & 1.905 & 1.153 & 2.501 & 3.813 & 4.308 & 4.519 & 5 \\
 9262 & 02:54:06.5 & $-$08:36:00 & 18.929 & 1.859 & 1.118 & 2.349 & 3.554 & 3.993 & 4.194 & 2 \\
11568 & 03:43:49.5 & +72:27:16 & 19.004 & 1.233 & 0.731 & 1.437 & 3.858 & 4.154 & 3.776 & 2 \\
12010 & 03:50:43.8 & $-$28:41:13 & 19.046 & 1.845 & 1.191 & 2.604 & 3.942 & 4.443 & 4.676 & 2 \\
12291 & 03:59:20.1 & +16:03:06 & 18.869 & 1.870 & 1.178 & 2.343 & 3.661 & 4.185 & 4.685 & 2 \\
12754 & 04:16:31.5 & +75:24:04 & 18.808 & 1.468 & 0.943 & 1.826 & 3.004 & 3.701 & 3.597 & 2 \\
12790 & 04:13:26.2 & $-$10:43:34 & 17.883 & 0.602 & 0.295 & 0.528 & 2.054 & 2.456 & 2.276 & 3 \\
12852 & 04:15:44.6 & +28:15:09 & 19.054 & 1.801 & 1.285 & 2.848 & 4.205 & 4.818 & 5.101 & 2 \\
13056 & 04:21:06.0 & +18:55:30 & 19.155 & 1.836 & 1.361 & 2.948 & 4.440 & 5.070 & 5.393 & 2 \\
13487 & 04:33:03.6 & +14:50:51 & 18.699 & 1.972 & 0.178 & 2.266 & 3.669 & 4.275 & 4.660 & 1 \\
14601 & 05:10:39.1 & +19:24:07 & 18.526 & 1.104 & 0.695 & 1.364 & 3.895 & 4.304 & 4.449 & 2 \\
15844 & 05:56:20.0 & $-$22:46:14 & 19.401 & 1.671 & 1.211 & 2.659 & 3.955 & 4.414 & 4.634 & 1 \\
15954 & 06:01:43.9 & $-$18:10:42 & 19.183 & 1.754 & 1.210 & 2.651 & 3.977 & 4.449 & 4.816 & 1 \\
16693 & 06:35:07.7 & +35:58:22 & 18.724 & 1.589 & 1.163 & 2.488 & 3.584 & 4.290 & 4.322 & 1 \\
16807 & 06:40:36.3 & +49:59:43 & 19.577 & 1.568 & 1.310 & 2.792 & 4.217 & 4.834 & 4.868 & 1 \\
17301 & 07:01:57.1 & +30:15:08 & 17.796 & 1.619 & 1.187 & 2.685 & 4.020 & 4.542 & 4.798 & 1 \\
17374 & 07:05:44.8 & +51:41:31 & 18.254 & 0.879 & 0.507 & 0.946 & 2.803 & 3.109 & 3.162 & 1 \\
19204 & 08:15:37.3 & $-$00:02:53 & 18.340 & 2.063 & 1.380 & 2.882 & 4.413 & 4.944 & 5.141 & 1 \\
21105 & 09:10:24.4 & +27:19:20 & 19.630 & 1.891 & 1.342 & 2.912 & 4.252 & 4.778 & 5.222 & 1 \\
21227 & 09:12:45.5 & $-$01:40:58 & 19.785 & 1.843 & 1.394 & 3.173 & 4.692 & 5.214 & 5.633 & 3 \\
22366 & 09:41:48.1 & +17:21:46 & 18.431 & 1.707 & 1.140 & 2.441 & 3.705 & 4.174 & 4.438 & 1 \\
22965 & 09:56:04.0 & +17:59:45 & 19.376 & 1.785 & 1.176 & 2.559 & 3.932 & 4.470 & 4.647 & 2 \\
23883 & 10:17:15.6 & +46:54:06 & 19.519 & 1.661 & 1.284 & 2.709 & 4.019 & 4.519 & 4.481 & 2 \\
25133 & 10:42:46.9 & $-$18:01:02 & 18.327 & 1.770 & 1.167 & 2.602 & 3.867 & 4.461 & 4.611 & 3 \\
31444 & 12:39:41.7 & +35:18:25 & 18.528 & 1.806 & 1.128 & 2.823 & 4.130 & 4.647 & 4.920 & 2 \\
33309 & 13:13:46.5 & $-$28:58:30 & 18.592 & 1.681 & 1.112 & 2.339 & 3.508 & 3.969 & 4.341 & 2 \\
34372 & 13:31:28.2 & +24:47:11 & 19.608 & 1.884 & 1.241 & 2.851 & 4.217 & 4.815 & 5.072 & 2 \\
35216 & 13:47:21.7 & $-$26:53:45 & 18.205 & 1.830 & 1.110 & 2.457 & 3.675 & 4.237 & 4.435 & 3 \\
36953 & 14:19:10.4 & +20:23:30 & 18.662 & 2.086 & 1.276 & 2.801 & 4.143 & 4.663 & 4.948 & 2 \\
37223 & 14:24:37.5 & $-$18:53:36 & 18.850 & 1.670 & 1.030 & 2.103 & 3.332 & 4.019 & 4.132 & 3 \\
38350 & 14:47:01.0 & $-$15:00:20 & 18.846 & 1.890 & 1.087 & 2.277 & 3.505 & 4.045 & 4.448 & 4 \\
38947 & 14:58:08.5 & +39:38:51 & 18.559 & 1.733 & 1.198 & 2.707 & 3.965 & 4.449 & 4.756 & 2 \\
41222 & 15:48:26.1 & $-$19:53:60 & 18.131 & 1.948 & 1.708 & 3.756 & 3.493 & 4.034 & 4.132 & 5 \\
42233 & 16:10:59.3 & +56:11:36 & 19.368 & 1.719 & 1.078 & 2.444 & 3.730 & 4.309 & 4.378 & 1 \\
42815 & 16:26:24.6 & +28:56:26 & 19.041 & 1.637 & 1.169 & 2.601 & 3.935 & 4.527 & 4.641 & 4 \\
45501 & 17:48:04.3 & +46:23:54 & 17.982 & 1.773 & 1.191 & 2.637 & 3.899 & 4.418 & 4.703 & 4 \\
46775 & 18:37:09.5 & +52:20:31 & 20.025 & 1.478 & 1.103 & 2.334 & 4.328 & 4.898 & 5.352 & 4 \\
47221 & 18:56:42.7 & +34:13:59 & 19.266 & 1.822 & 1.185 & 2.687 & 4.061 & 4.528 & 4.859 & 4 \\
48482 & 19:57:34.6 & $-$11:29:59 & 18.928 & 1.723 & 1.122 & 2.284 & 3.411 & 3.997 & 4.342 & 3 \\
50025 & 20:52:09.6 & $-$23:18:06 & 18.600 & 1.965 & 1.922 & 4.120 & 2.368 & 2.577 & 2.922 & 2 \\
50476 & 21:05:14.0 & $-$24:46:52 & 17.768 & 1.734 & 1.291 & 2.939 & 4.409 & 4.901 & 5.150 & 2 \\
50931 & 21:17:50.4 & $-$29:22:05 & 17.820 & 1.771 & 1.133 & 2.506 & 3.795 & 4.362 & 4.603 & 2 \\
51256 & 21:26:03.5 & +14:29:18 & 19.549 & 1.688 & 1.268 & 2.908 & 4.351 & 4.854 & 5.178 & 4 \\
51509 & 21:32:49.3 & +12:50:36 & 17.941 & 1.669 & 1.293 & 2.906 & 4.374 & 4.925 & 5.250 & 3 \\
52449 & 21:55:48.0 & $-$11:21:43 & 17.402 & 1.947 & 1.191 & 2.527 & 3.753 & 4.215 & 4.490 & 3 \\
54542 & 22:40:46.2 & +26:08:20 & 18.877 & 1.719 & 1.131 & 2.448 & 3.682 & 4.367 & 4.422 & 3 \\
54950 & 22:48:58.6 & +32:08:26 & 18.984 & 1.866 & 1.713 & 3.759 & 3.958 & 4.399 & 4.212 & 2 \\
55103 & 22:52:45.5 & $-$25:25:21 & 18.411 & 1.811 & 1.260 & 2.753 & 4.093 & 4.678 & 4.997 & 2 \\
55262 & 22:55:08.5 & +25:51:14 & 19.103 & 1.618 & 1.112 & 2.329 & 3.628 & 4.309 & 4.585 & 2 \\

\enddata
\end{deluxetable}

\begin{deluxetable}{lccc}
\tablecolumns{4}
\tablewidth{0pc}
\tablecaption{
Line indices defined by \cite{gizis97}
\label{table:gizisind}
}
\tablehead{
\colhead{Line} & \colhead{Line Band} & \colhead{Sideband 1} & \colhead{Sideband 2} \\
\colhead{} & \colhead{(\AA)} & \colhead{(\AA)} & \colhead{(\AA)}
}
\startdata

CaH 2 & 6814.0 - 6846.0 & 7042.0 - 7046.0 & \\
CaH 3 & 6960.0 - 6990.0 & 7042.0 - 7046.0 & \\
TiO 5 & 7126.0 - 7135.0 & 7042.0 - 7046.0 & \\

\enddata
\end{deluxetable}

\begin{deluxetable}{llllllllc}
\tablecolumns{9}
\tablewidth{0pc}
\tablecaption{
Measured line indices and spectral classifications
\label{table:gizis}
}
\tablehead{
\colhead{NLTT} &  \colhead{$V$} & \colhead{$V-J$} & \colhead{Date} & \colhead{Telescope}
& \colhead{CaH2} & \colhead{CaH3} & \colhead{TiO5} & \colhead{Class}
}
\startdata

1352 & 19.053 & 4.386 & 04 Oct 2003 & CTIO & 0.296 & 0.450 & 0.354 & sdM6.0 \\
4308 & 19.18 & 4.164 & 28 Oct 2003 & MDM & 0.344 & 0.622 & 0.366 & M4.5 \\
4697 & 19.594 & 4.873 & 09 Nov 2003 & MDM & 0.347 & 0.608 & 0.246 & M5.0 \\
5282 & 19.136 & 3.947 & 27 Jul 2003 & CTIO & 0.128 & 0.049 & 0.740 & esdM9.5 \\
5399 & 19.903 & 4.48 & 03 Oct 2003 & CTIO & 0.014 & 0.737 & 0.130 & sdM6.0 \\
6590 & 18.947 & 3.813 & 08 Oct 2004 & MDM & 0.308 & 0.474 & 0.577 & sdM5.5 \\
9262 & 18.929 & 3.554 & 03 Oct 2003 & CTIO & 0.420 & 0.610 & 0.798 & esdM3.5 \\
11568 & 19.004 & 3.858 & 07 Oct 2004 & MDM & 0.934 & 0.886 & 1.112 & esdK7 \\
12010 & 19.046 & 3.942 & 04 Oct 2003 & CTIO & 0.453 & 0.619 & 0.650 & sdM3.5 \\
12291 & 18.869 & 3.661 & 08 Oct 2004 & MDM & 0.510 & 0.610 & 0.775 & sdM3.0 \\
12790 & 17.883 & 2.054 & 05 Oct 2003 & CTIO & 1.026 & 0.992 & 0.987 & K5 \\
12852 & 19.054 & 4.205 & 07 Oct 2004 & MDM & 0.443 & 0.666 & 0.475 & M3.5 \\
15844 & 19.401 & 3.955 & 03 Oct 2003 & CTIO & 0.313 & 0.477 & 0.499 & sdM5.5 \\
15954 & 19.183 & 3.977 & 04 Oct 2003 & CTIO & 0.267 & 0.412 & 0.441 & sdM6.5 \\
16693 & 18.724 & 3.584 & 04 Nov 2004 & MDM & 0.435 & 0.689 & 0.491 & M3.0 \\
17301 & 17.796 & 4.02 & 28 Oct 2003 & MDM & 0.421 & 0.668 & 0.422 & M3.5 \\
19204 & 18.34 & 4.413 & 15 Apr 2004 & MDM & 0.403 & 0.648 & 0.508 & sdM3.5 \\
22366 & 18.431 & 3.705 & 14 Apr 2004 & MDM & 0.373 & 0.813 & 0.678 & sdM2.5 \\
23883 & 19.519 & 4.019 & 15 Apr 2004 & MDM & 0.464 & 0.699 & 0.572 & M3.0 \\
25133 & 18.327 & 3.867 & 11 Jun 2004 & CTIO & 0.406 & 0.605 & 0.665 & sdM4.0 \\
31444 & 18.528 & 4.13 & 14 Apr 2004 & MDM & 0.378 & 0.579 & 0.468 & sdM4.0 \\
33309 & 18.592 & 3.508 & 11 Jun 2004 & CTIO & 0.492 & 0.693 & 0.751 & sdM2.5 \\
34372 & 19.608 & 4.217 & 15 Apr 2004 & MDM & 0.215 & 0.488 & 0.413 & sdM6.5 \\
35216 & 18.205 & 3.675 & 10 Jun 2004 & CTIO & 0.442 & 0.701 & 0.652 & sdM3.0 \\
36953 & 18.662 & 4.143 & 09 Jun 2003 & MDM & 1.132 & 1.064 & 1.002 & K \\
37223 & 18.85 & 3.332 & 28 Jul 2003 & CTIO & 0.225 & 0.456 & 0.606 & esdM6.0 \\
 &  &  & 10 Jun 2004 & CTIO & 0.319 & 0.668 & 0.810 & esdM3.5 \\
38350 & 18.846 & 3.505 & 11 Jun 2004 & CTIO & 0.457 & 0.759 & 0.801 & sdM2.5 \\
38947 & 18.559 & 3.965 & 15 Jun 2003 & MDM & 0.452 & 0.694 & 0.541 & M3.0 \\
42233 & 19.368 & 3.73 & 14 Apr 2004 & MDM & 0.354 & 0.723 & 0.403 & M3.5 \\
42815 & 19.041 & 3.935 & 09 Jun 2003 & MDM & 1.084 & 1.026 & 0.973 & K5 \\
45501 & 17.982 & 3.899 & 09 Jun 2003 & MDM & 0.391 & 0.630 & 0.513 & sdM4.0 \\
47221 & 19.266 & 4.061 & 14 Apr 2004 & MDM & 0.855 & 1.230 & 1.273 & esdK \\
48482 & 18.928 & 3.411 & 26 Jul 2003 & CTIO & 0.210 & 0.502 & 0.368 & sdM6.0 \\
50476 & 17.768 & 4.409 & 27 Jul 2003 & CTIO & 0.346 & 0.599 & 0.387 & sdM4.5 \\
 &  &  & 28 Jul 2003 & CTIO & 0.379 & 0.625 & 0.402 & M4.0 \\
 &  &  & 03 Oct 2003 & CTIO & 0.356 & 0.596 & 0.393 & sdM4.5 \\
50931 & 17.82 & 3.795 & 26 Jul 2003 & CTIO & 0.385 & 0.604 & 0.512 & sdM4.0 \\
51256 & 19.549 & 4.351 & 09 Nov 2003 & MDM & 0.491 & 0.659 & 0.385 & M3.5 \\
51509 & 17.941 & 4.374 & 28 Oct 2003 & MDM & 0.388 & 0.644 & 0.388 & M4.0 \\
52449 & 17.402 & 3.753 & 26 Jul 2003 & CTIO & 0.322 & 0.456 & 0.691 & esdM5.0 \\
54542 & 18.877 & 3.682 & 28 Oct 2003 & MDM & 0.404 & 0.629 & 0.611 & sdM3.5 \\
54950 & 18.984 & 3.958 & 09 Nov 2003 & MDM & 1.045 & 0.980 & 0.977 & K5 \\
55103 & 18.411 & 4.093 & 26 Jul 2003 & CTIO & 0.306 & 0.537 & 0.297 & sdM5.0 \\
 &  &  & 27 Jul 2003 & CTIO & 0.299 & 0.609 & 0.349 & sdM4.5 \\
55262 & 19.103 & 3.628 & 07 Oct 2004 & MDM & 1.003 & 0.957 & 0.929 & K6 \\

\enddata
\end{deluxetable}


\begin{thebibliography}{}


\bibitem[Gizis(1997)]{gizis97}
Gizis, J. E. 1997, \aj, 113, 806


\bibitem[Gizis \& Reid(1997)]{gr97}
Gizis, J. E., \& Reid, I. N. 1997, \pasp, 109, 849

\bibitem[Gliese \& Jahreiss(1991)]{pCNS3}
Gliese, W., \& Jahreiss, H. 1991.  Preliminary Version of the Third Catalog of Nearby Stars.

\bibitem[Gould(2003)]{g4588}
Gould, A. 2003, \apj, 583, 765

\bibitem[Gould \& Salim(2003)]{gs03}
Gould, A., \& Salim, S. 2003, \apj, 582, 1001

\bibitem[Hawley et al.(1996)]{pmsu2}
Hawley, S. L., Gizis, J. E., \& Reid, I. N. 1996, \aj, 112, 2799


\bibitem[Lepine et al.(2003)]{lrs2003}
Lepine, S., Rich, R. M., \& Shara, M. M. 2003, \aj, 125, 1598

\bibitem[Lepine et al.(2007)]{lrs2007}
Lepine, S., Rich, R. M., \& Shara, M. M. 2007, \apj, in press.

\bibitem[Lepine \& Shara(2005)]{lspm}
Lepine, S., \& Shara, M. M. 2005, \aj, 129, 1483

\bibitem[Marshall(2007)]{paper1}
Marshall, J. L. 2007, \aj, 134, 778

\bibitem[Monet(1996)]{monet96}
Monet, D. G. 1996, \baas, 188, 5405

\bibitem[Monet(1998)]{monet98}
Monet, D. G. 1998, \baas, 193, 112003

\bibitem[Reid \& Gizis(2005)]{rg05}
Reid, I. N., \& Gizis, J. E. 2005, \pasp, 117, 676

\bibitem[Reid et al.(1995)]{pmsu1}
Reid, I. N., Hawley, S. L., Gizis, J. E. 1995, \aj, 110, 1838

\bibitem[Salim \& Gould(2002)]{sg02}
Salim, S., \& Gould, A. 2002, \apj, 575, 83

\bibitem[Salim \& Gould(2003)]{sg03}
Salim, S., \& Gould, A. 2003, \apj, 582, 1011

\bibitem[Skrutskie et al.(1997)]{2mass}
Skrutskie, M. F., et al. 1997, in The Impact of Large-Scale Near-IR Sky Survey, ed. F. Garzon et al. (Dordrecht: Kluwer), 187

%


\end{thebibliography}
\end{document}